\documentclass[12pt]{iopart}
\usepackage{epsfig}

\def\mean#1{\left<{#1}\right>}
\def\Journal#1#2#3#4{{#1}{\bf #2}, #3 (#4)}

\def\JPG{{J. Phys}~{\bf G}}

\def\NPB{{Nucl. Phys.}~{\bf B}}
\def\PLB{{Phys. Lett.}~{\bf B}}
\def\PR{Phys. Rep.\ }
\def\PS{Physica. Scripta.\ }
\def\PRL{Phys. Rev. Lett.\ }
\def\PRD{{Phys. Rev.}~{\bf D}}

\begin{document}

\title[Probing jet properties via two particle correlation method]{Probing jet properties via two particle correlation method}

\author{Jiangyong Jia\dag\
\footnote[3]{jjia@shang.nevis.columbia.edu} }

\address{\dag\ Columbia University, New York, NY 10027 and Nevis Laboratories, Irvington, NY 10533, USA}

\begin{abstract}
The formulae for calculating jet fragmentation momentum,
$\left<j_T^2\right>$, and parton transverse momentum,
$\left<k_T^2\right>$, and conditional yield are discussed in two
particle correlation framework. Additional corrections are derived
to account for the limited detector acceptance and inefficiency,
for cases when the event mixing technique is used. The validity of
our approach is confirmed with Monte-carlo simulation.
\end{abstract}



%\maketitle

\section{Introduction}

In pp, pA or AA collisions at RHIC, the high $p_T$ part of the
hadron spectra is dominated by the two-body hard-scattering
process. In this process, the two scattered partons typically
appear as a pair of almost back-to-back jets. The properties of
the di-jet system can be characterized by the jet fragmentation
momentum, $j_T$, the parton transverse momentum, $k_T$, and the
fragmentation function, $D(z)$. Specifically, $j_T$, $k_T$ and
$D(z)$ describe the spread of hadrons around the jet axis, the
relative orentation of the back-to-back jets and the jet
multiplicity, respectively.

Traditionally, energetic jets were reconstructed directly using
standard jet finding algorithms\cite{conemethod,ktmethod}. In
heavy-ion collisions, due to the large amount of soft background,
direct jet reconstruction is difficult. Even in pA or pp
collisions, the range of energy accessible to direct jet
reconstruction is probably limited to $p_T>10$ GeV~\cite{star},
below which the `underlying event' background contamination become
important. The situation is even more complicated for finite
acceptance detectors like PHENIX due to the leakage of the jet
cone outside the acceptance.

The two-particle-correlation technique provides an alternative way
to access the properties of the jet. It is based on the fact that
the fragments are tightly correlated in azimuth $\phi$ and
pseudo-rapidity $\eta$ and the jet signal manifests itself as a
narrow peak in $\Delta\phi$ and $\Delta\eta$ space. The jet
properties are extracted statistically by accumulating many
events. This method was initially used in 70's in searching for
jet signals in pp collisions at CERN ISR~\cite{ccor,ccor1,cchk}.
It overcomes problems with the underlying event background, probes
the jet signal at lower $p_T$, and has recently excited renewed
interest at RHIC~\cite{jetexp,jetthe,rak}.

Experimentally, jet measurement is challenging due to detector
inefficiency and limited experimental acceptance. A certain
fraction of jet pairs is lost either because the track is not
found or because part of the jet cone falls outside the
acceptance. The average jet pair detection efficiency can be
estimated statistically using event mixing technique. The main
goal of the paper is to establish the procedures for measuring the
jet shape and jet multiplicity independent of the detector
efficiency and acceptance.

The discussion is split into three sections. In
Section.\ref{sec:formula}, we lay out the formulae for extracting
$j_T$, $k_T$ and conditional yield. In the interest of space, the
reader is encouraged to see Ref.\cite{rak,feynman} for more
details on the definition of the variables. In
Section.\ref{sec:acceptance}, we discuss the event mixing
technique necessary in correcting for limited acceptance and
inefficiency, where we use the PHENIX detector as an example. We
discuss separately the two dimensional ($\Delta\eta$,
$\Delta\phi$) and one dimensional ($\Delta\phi$) correlation and
derive the normalization factors for the conditional yield in each
case. In Section.\ref{sec:simulation} we verify the procedure of
correcting for finite acceptance with Pythia simulation.
\section{Some formulae}
\label{sec:formula}
\subsection{Formulae for $j_T$, $k_T$}
\begin{figure}[ht]
\begin{center}
\epsfig{file=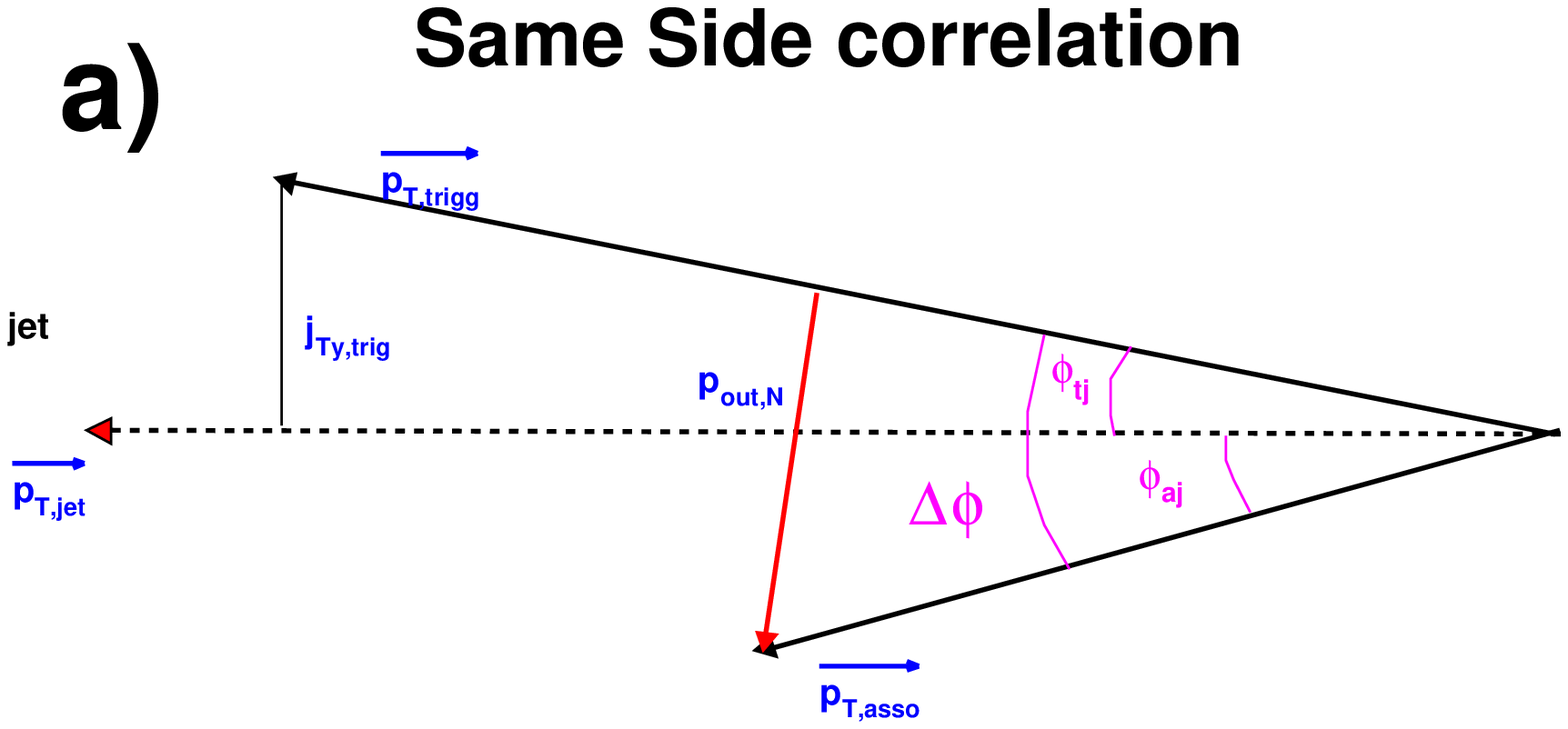,width=0.7\linewidth}\\
\epsfig{file=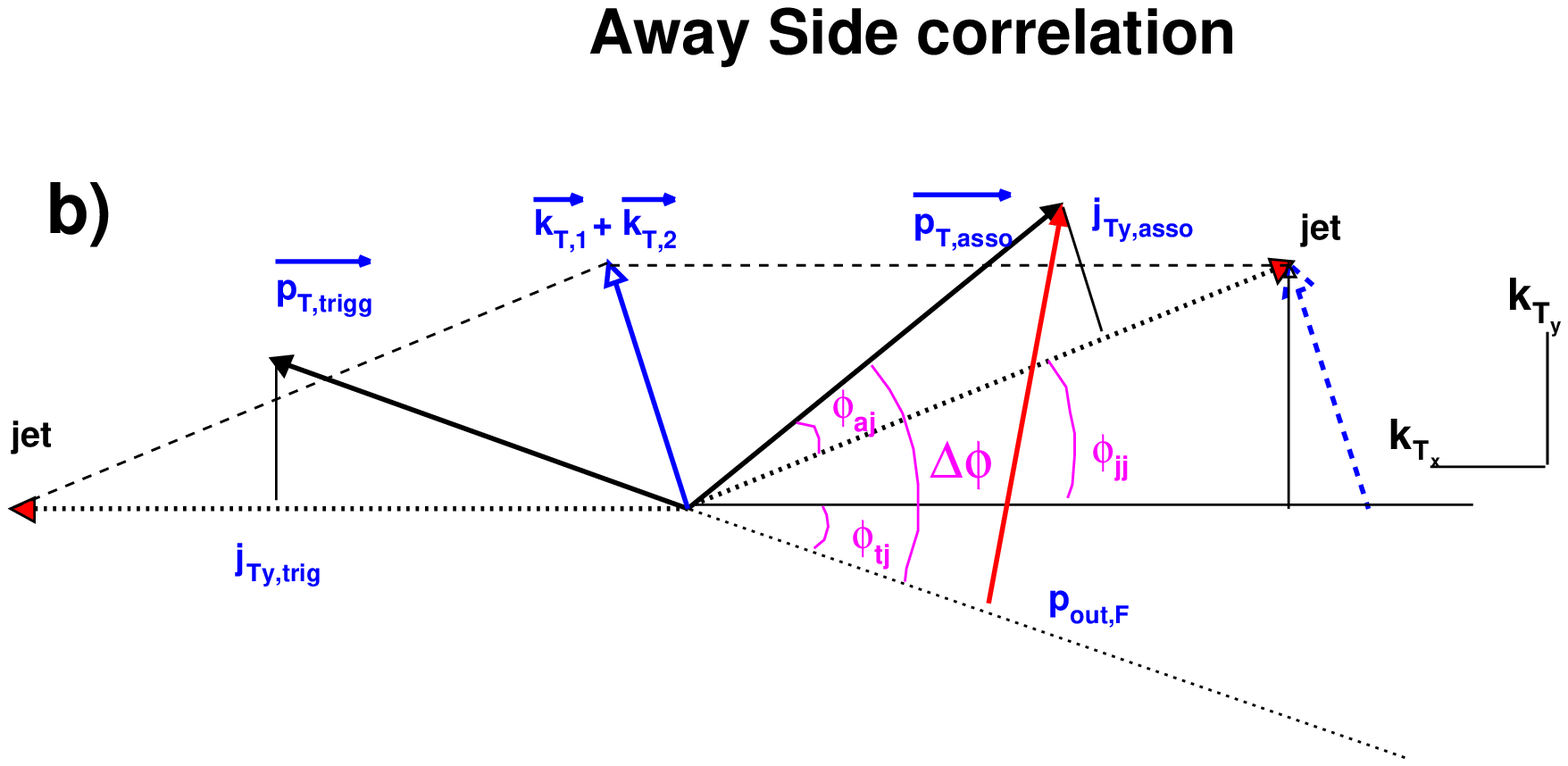,width=0.7\linewidth}
\caption{\label{fig:frag} Schematic view of jet fragmentation. a)
same side jet, b) away side jet.}
\end{center}
\end{figure}

Fig.\ref{fig:frag} illustrates the single jet fragmentation (left
panel) and away side jet fragmentation (right panel). The jet
fragmentation momentum, $j_T$, and parton initial momentum, $k_T$,
determine the relative orientation of the fragmented hadrons.
$j_T$ and $k_T$ are vectors, their projection in the azimuthal
plane perpendicular to the jet direction are denoted as $j_{T_y}$
and $k_{T_y}$. For single jet fragmentation, if we denote
$\Delta\phi$, $\phi_{tj}$, and $\phi_{aj}$ as the angles between
trigger-associated, trigger-jet and associated-jet, respectively,
then the following relations are true:
\begin{eqnarray}
\label{eq:sinnear}\nonumber
\Delta\phi= \phi_{tj} + \phi_{aj}\quad&,&\quad\sin(\Delta\phi) = \frac{p_{out,N}}{p_{T,asso}}\quad,\\
\sin(\phi_{tj}) =
\frac{j_{T_y,trig}}{p_{T,trig}}\equiv x_{j,trig}\quad&,&\quad
\sin(\phi_{aj}) = \frac{j_{T_y,asso}}{p_{T,asso}}\equiv
x_{j,asso}\quad.
\end{eqnarray}
Here $p_{out,N}$ is the component of associated particle $p_{T}$
($p_{T,asso}$) perpendicular to the trigger particle $p_T$
($p_{T,trig}$). Assuming $\phi_{tj}$ and $\phi_{aj}$ are
statistically independent, we have (cross terms average to 0),
\begin{eqnarray}
\mean{\sin^2\Delta\phi}= \mean{\sin^2\phi_{tj}\cos^2\phi_{aj}} +
\mean{\sin^2\phi_{aj}\cos^2\phi_{tj}} \label{eq:jt1}
\end{eqnarray}

Similarly for the far-side correlation, we have
\begin{eqnarray}
\label{eq:sinfar} \Delta\phi &=& \phi_{tj} + \phi_{aj} +
\phi_{jj}\quad,\quad \sin(\Delta\phi) = \frac{p_{out,F}}{p_{T,asso}}\quad,
\end{eqnarray}
where $\phi_{jj}$ represents the angle between the two jets.
Expanding $\sin^2\Delta\phi$ and dropping all cross terms (which
average to 0), we get
\begin{eqnarray}
\label{eq:kt1} \nonumber\mean{\sin^2\Delta\phi} &=&
\mean{\left(\sin\phi_{tj}\cos\phi_{aj}cos\phi_{jj}\right)^2}+
\mean{\left(\sin\phi_{aj}\cos\phi_{tj}cos\phi_{jj}\right)^2}+\\
&&\mean{\left(\sin\phi_{jj}\cos\phi_{aj}cos\phi_{tj}\right)^2}+
\mean{\left(\sin\phi_{tj}\sin\phi_{aj}sin\phi_{jj}\right)^2}
\end{eqnarray}

Let's define $z_{trig}= p_{T,trig}/p_{T,jet}$ and $x_k =
\sqrt{2}k_{T_y}/p_{T,jet}=\sqrt{2}k_{T_y}z_{trig}/p_{T,trig}$. If
$\phi_{jj}$ is small, or $p_{T,jet}\gg k_{T_y}$, then $\sin(\phi_{jj} ) \approx \left(k_{T_y ,trig}  +
k_{T_y ,asso}\right)/p_{T,jet}$, $i.e.$ $k_T$ only
affects the jet direction. Thus statistically we obtain
\begin{eqnarray}
\label{eq:kt2}\mean{sin^2(\phi_{jj})}\approx
2\mean{\frac{k^2_{T_y}}{p^2_{T,jet}}}=\mean{x_{k}^2}\quad.
\end{eqnarray}
Substituting the $\sin$ and $\cos$ terms from Eq.~\ref{eq:sinnear}
and \ref{eq:sinfar} into Eq.~\ref{eq:jt1} and Eq.~\ref{eq:kt1}, we
obtain the equations for the RMS value of $j_{T_y}$ and $k_{T_y}$
(for a given variable $x$, $(x)_{RMS}\equiv\sqrt{\mean{x^2}}$),
\begin{eqnarray}
\label{eq:jt2} \left( {j_{T_y } } \right)_{RMS}  = \sqrt
{{\left\langle {p_{out,N}^2 } \right\rangle } \mathord{\left/
 {\vphantom {{\left\langle{p_{out,N}^2 }\right\rangle}{1 + \left\langle {x_h^2 } \right\rangle  - 2\left\langle
{x_{j,trig}^2 } \right\rangle}}} \right.
 \kern-\nulldelimiterspace} {\left(1 + \left\langle {x_h^2 } \right\rangle  - 2\left\langle
{x_{j,trig}^2 } \right\rangle\right)}}
\end{eqnarray}
\begin{eqnarray} \label{eq:kt3}
\left(k_{T_y}z_{trig}\right)_{RMS} =
\sqrt{\frac{{\left\langle {p_{out,F}^2 } \right\rangle  -
\left\langle {p_{out,N}^2 } \right\rangle \left( {1 - \left\langle
{x_k^2 } \right\rangle } \right)}}{2\left(\left\langle {x_h^2 }
\right\rangle - \left\langle {x_h^2 x_{j,trig}^2 } \right\rangle -
\left\langle {x_{j,trig}^2 } \right\rangle +2\left\langle
{x_{j,trig}^4}\right\rangle\right)}}
\end{eqnarray}
where $x_h = p_{T,asso}/p_{T,trig}$. Assuming the $\Delta\phi$
follows gaussian statistics, a simple Taylor expansion connects
$p_{out}$ with the jet width, $\sigma$:
\begin{eqnarray}
\label{eq:expan}\nonumber \mean{p_{out}^2} &=&
\mean{p_{T,asso}^2sin^2\Delta\phi}\\\nonumber
&\approx&\mean{p_{T,asso}^2}[sin\mean{\Delta\phi^2}-\frac{\mean{\Delta\phi^4}}{3}]\\
&=&\mean{p_{T,asso}^2}[sin\sigma^2-\sigma^4]
\end{eqnarray}
By replacing the $p_{out}$ terms in Eq.\ref{eq:jt1} and
Eq.\ref{eq:kt1}, we can derive the relation between jet width and RMS value
of $j_{T_y}$ and $k_{T_y}$.

Since Eq.~\ref{eq:jt2} and Eq.~\ref{eq:kt3} contain variables
$x_{j,trig}$ and $x_k$ that depend on $j_{T_y}$ and $k_{T_y}$, we
have to calculate $(j_{T_y})_{RMS}$ and $(k_{T_y})_{RMS}$
iteratively. However when trigger and associated particle $p_T$
are much larger than typical $j_T$ value, the near side jet width
$\sigma_{N}$ is small and $x_{j,trig}\approx 0$. Hence
Eq.~\ref{eq:jt2} can be simplified to,
\begin{eqnarray}
\label{eq:jt3} \left(j_{T_y}\right)_{RMS} & \simeq &
\sqrt{\frac{sin\sigma^2_N\mean{p^2_{T,asso}}}{1+\mean{x_h^2}}}
\end{eqnarray}
For the far-side correlation, if the trigger $p_T$ is much larger than $k_{T_y}$
and $j_{T_y}$, then $\mean{x_k^2}\approx 0$,
$\mean{x_j^2}\approx 0$ and Eq.~\ref{eq:kt3} reduces to

\begin{eqnarray}
\label{eq:kt4} \left({k_{Ty} z_{trig}}\right)_{RMS} &\simeq&
\sqrt{\frac{\mean{p_{out,F}^2}-\mean{p_{out,N}^2}}{2\mean{x_h^2}}}\\\nonumber
&\simeq &\frac{1}{\sqrt{2\mean{x_h^2}}} \sqrt{\langle
p_{T,assoc}\rangle^2 \sin^2\sigma_F -
(1+\mean{x_h^2})\left(j_{T_y}\right)^2_{RMS}}
\end{eqnarray}

\begin{figure}[ht]
\begin{tabular}{c}
\begin{minipage}{1.0\linewidth}
\epsfig{file=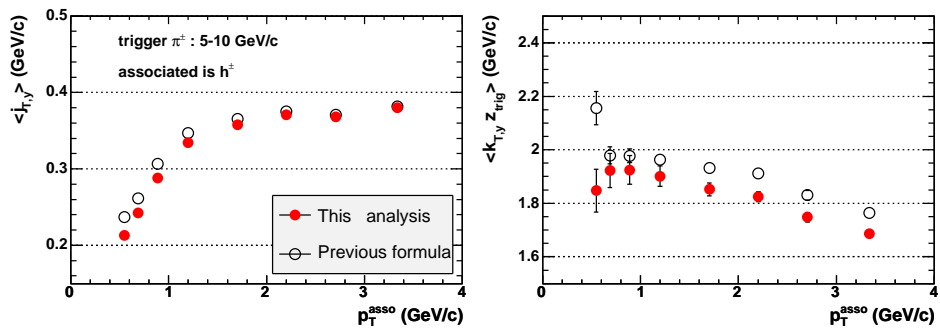,width=0.8\linewidth}
\end{minipage}
\\
\hspace*{-3.0cm}
\begin{minipage}{1.1\linewidth}
\caption{\label{fig:frag1} The comparison of the two sets of
formulae for $j_{T_y}$ RMS value (left panel) and
$z_{trig}k_{T_y}$ RMS value (right panel). Values are calculated
from jet width given by Pythia simulation.}
\end{minipage}
\end{tabular}
\end{figure}

Previously, Ref~\cite{rak} has derived the approximate relations
between the $j_T$, $k_T$ and the measured jet width in the near
side ($\sigma_{N}$) and the away side ($\sigma_{A}$) ($\mean{|a|}$
is the mean of the $|a|$, $\mean{|a|}^2 = 2/\pi \mean{a^2}$):
\small{\begin{eqnarray} \label{eq:jt4} \mean{|j_{T_y}|}  &=&
\sqrt{2/\pi}{{\left\langle {p_{T,asso} } \right\rangle \sigma _N }
\mathord{\left/
 {\vphantom {{\left\langle {p_T } \right\rangle \sigma _N } {\sqrt {1 + \left\langle {x_h } \right\rangle ^2 } }}} \right.
 \kern-\nulldelimiterspace} {\sqrt {1 + \left\langle {x_h } \right\rangle ^2 }
 }}\\
\label{eq:kt5} \mean{|k_{T_y}z_{trig}|} &=&
\frac{1}{\sqrt{2}\left\langle {x_h }
\right\rangle}\sqrt{\mean{p_{T,asso}}^2\sin^2\sqrt{\frac{2}{\pi}}\sigma_A-(1+\left\langle
{x_h } \right\rangle^2)\mean{|j_{T_y}|}^2}
\end{eqnarray}}\normalsize
One can see that Eq.\ref{eq:jt4} and Eq.\ref{eq:kt5} are very
close to our approximation Eq.\ref{eq:jt3} and Eq.\ref{eq:kt4}. To
quantify the difference between Eq.\ref{eq:jt2}-\ref{eq:kt3} and
Eq.\ref{eq:jt4}-\ref{eq:kt5}, we performed the correlation
analysis with $\pi^{\pm}-h^{\pm}$ correlation from the Pythia
event generator~\cite{pythia} and compared the $j_T$ $k_T$ value
calcualted from the two sets of formulae. The results are shown in
Fig.\ref{fig:frag1}, there are good agreements for $j_{T_y}$,
except at low $p_{T,asso}$ where the $\sigma_{N}$ become big and
Eq.\ref{eq:expan} has to be used. On the other hand, the
$\left<z_{trig}k_{T_y}\right>$ value from this analysis is 10\%
lower than the one calculated by Eq.\ref{eq:kt3}. The difference
could be caused by small angle approximation used in
Eq.\ref{eq:kt3}, but may also due to the difference between RMS
and mean: for given variable $v$ in a finite range,
$\sqrt{\left<v^2\right>} \geq \left<v\right>$. There is a dropping
trend in the $\left<z_{trig}k_{T_y}\right>$ value as function of
$p_{T,asso}$ in both cases. This trend can be explained by the
trigger bias effects~\cite{rakthis}, which we briefly touch on in
the next section.

\subsection{Formula for yield}

The jet fragmentation function, $D(z)$ or $D(p_T)$, represents the
associated yield per jet,
\begin{eqnarray}
D(z) &=& 1/N_{jet} dN_{asso}/dz, \qquad or \qquad D(p_T) =
1/N_{jet}dN_{asso}/dp_T
\end{eqnarray}
where $D(p_T) = D(z)/p_{T,jet}$, when the jet energy $p_{T,jet}$
is fixed.

If the two particles come from the same jet, the two particle
multiplicity distribution can be described by the di-hadron
fragmentation function,
\begin{eqnarray} \label{eq:cond1} D(z_1 ,z_2 ) =
1/N_{jet}d^2 N_h/dz_1 dz_2\quad .
\end{eqnarray}
The conditional fragmentation function, where the $p_T$ of one
particle is fixed, can be expressed by,
\begin{eqnarray}
\label{eq:cond2} \frac{{D(z_1 ,z_2 )}}{{D(z_1 )}} =
{{\frac{1}{{N_{jet} }}\frac{{d^2 N_h }}{{dz_1 dz_2 }}}
\mathord{\left/
 {\vphantom {{\frac{1}{{N_{jet} }}\frac{{d^2 N_h }}{{dz_1 dz_2 }}} {\frac{1}{{N_{jet} }}\frac{{dN_h }}{{dz_1 }}}}} \right.
 \kern-\nulldelimiterspace} {\frac{1}{{N_{jet} }}\frac{{dN_h }}{{dz_1 }}}}\quad .
\end{eqnarray}

Experimentally, we measure jet properties from two particle
azimuthal correlation. The jet signal typically appears as two
distinct peaks in $\Delta\phi$, i.e. at $\Delta\phi=0$ for same
jet and at $\Delta\phi=\pi$ for away side jet. By fitting the
peaks with gaussian function, we extract the associated hadron
yield per trigger or the conditional yield ($CY$),
\begin{eqnarray}
\label{eq:cond3} CY(\Delta\phi,\Delta\eta) = \frac{1}{N_{trig}}
\frac{d^2N}{d\Delta\phi d\Delta\eta}
\end{eqnarray}
Where $N_{trig}$ is the number of trigger particles, and
$\frac{d^2N}{d\Delta\phi d\Delta\eta}$ is the number of jet pairs
in a fixed $\Delta\phi$ and $\Delta\eta$ bin. Since
$N_{trig}\propto D(z_{trig})$ and $\frac{d^2N}{d\Delta\phi
d\Delta\eta} \propto D(z_{trig},z_{asso})$, the $CY$ integrated
over the same jet (around $\Delta\phi=0$) directly corresponds to
the conditional fragmentation function. Unfortunately, the
fragmentation variable $z$ is not directly accessible in the
correlation method. Instead, $CY$ is often expressed as function
of a different variable $x_E$~\cite{ccor}, defined as $x_E =
\overrightarrow{p_T}\cdot\overrightarrow{p_{T,trig}}/|p_{T,trig}|^2=
z_{asso}/z_{trig}$.

The di-hadron fragmentation function (Eq.\ref{eq:cond1}),
conditional fragmentation function (Eq.\ref{eq:cond2}) and
conditional yield (Eq.\ref{eq:cond3}) are defined for the near
side correlation for which the two particles belong to the same
jet. They can be extended to describe the correlation of the two
particles from the back-to-back jets. In this case,
Eq.\ref{eq:cond2} represents the hadron-triggered fragmentation
function similar to the one used in~\cite{wang}. In a naive
parton-parton scattering picture and for fixed jet energy, the
fragmentation of the away side jet and triggering jet can be
factorized, i.e.
 $D(z_1 ,z_2 ) = \frac{1}{{N_{jet}
}}\frac{{dN_{trig}}}{dz_1} \frac{1}{{N_{jet}
}}\frac{{dN_{asso}}}{dz_2}$. Thus the $CY$ is related to the
unbiased fragmentation function by a scale factor,
$\mean{z_{trig}}$,
\begin{eqnarray}
\label{eq:cond4} CY(x_E ) &=& \frac{1}{{N_{trig}
}}\frac{{dN_h}}{{dx_E }} = z_{trig} D(z)\quad.
\end{eqnarray}
$\mean{z_{trig}}$ was found to be around 0.75-0.95 at
ISR~\cite{ccor1} for high $p_T$ leading pions (4-12 GeV/c) and
scales with $x_T$ in $\sqrt{s}$~\cite{ccor1}.

In reality, due to the intrinsic $k_T$ or radiative corrections,
the independent fragmentation assumption for parton-parton
scattering is not strictly valid. In addition, for given given
$p_{T,trig}$, the original jet energy is not fixed, but depends on
the $p_{T,asso}$, which implies that $z_{trig}$ also depends on
$p_{T,asso}$. This can be easily understood from the fact that the
fractional momentum can't exceed 1: $0<z<1$, while $x_E$ is not
bounded from above: $0<x_E<\infty$. Simple simulation indicates
that $z_{trig}$ is relatively stable when $p_{T,asso}<p_{T,trig}$,
but quickly decrease when $p_{T,asso}$ is larger than
$p_{T,trig}$~\cite{rakthis}. In summary, these bias effects,
caused by the requirement of a trigger hadron on the opposite
side, makes Eq.\ref{eq:cond4} at best an approximation. The
trigger bias effect is also responsible for the decreasing of
$\left<z_{trig}k_{T_y}\right>$ seen in Fig.\ref{fig:frag1}.

\section{Corrections for limited acceptance and efficiency}
\label{sec:acceptance}
\subsection{$CY$ in $\Delta\phi$ and $\Delta\eta$}
For a detector with limited aperture, a fraction of the jet cone
falls outside the acceptance. The fractional loss depends on the
jet direction: the loss for jet pointing to the corner of the
detector is larger than jet pointing to the center of the
detector. Assuming the jet production rate is uniform in azimuth
direction $\phi$ and pseudo-rapidity $\eta$, we can construct an
average pair acceptance function using event mixing technique (One
trigger particle is randomly combined with an associated particle
from a different event.)
\begin{eqnarray}
\label{eq:acc1} {{d^2N^{mix} } \mathord{\left/
 {\vphantom {{d^2N^{mix} } {d\Delta \phi d\Delta \eta }}} \right.
 \kern-\nulldelimiterspace} {d\Delta \phi d\Delta \eta }}
=Acc(\Delta\phi,\Delta\eta)m_0
\end{eqnarray}
$m_0$ represents the background level when the associated particle
is not constrained. Given the limited PHENIX $\eta$ coverage, it
is safe to assume that $m_0$ is constant~\cite{pythiaeta}.
$Acc(\Delta\phi,\Delta\eta)$ is the pair acceptance function,
which represents the probability of detecting the associated
particle when the trigger particle is detected.

\begin{figure}[ht]
\begin{tabular}{rl}
\begin{minipage}{0.6\linewidth}
\begin{flushright}
\epsfig{file=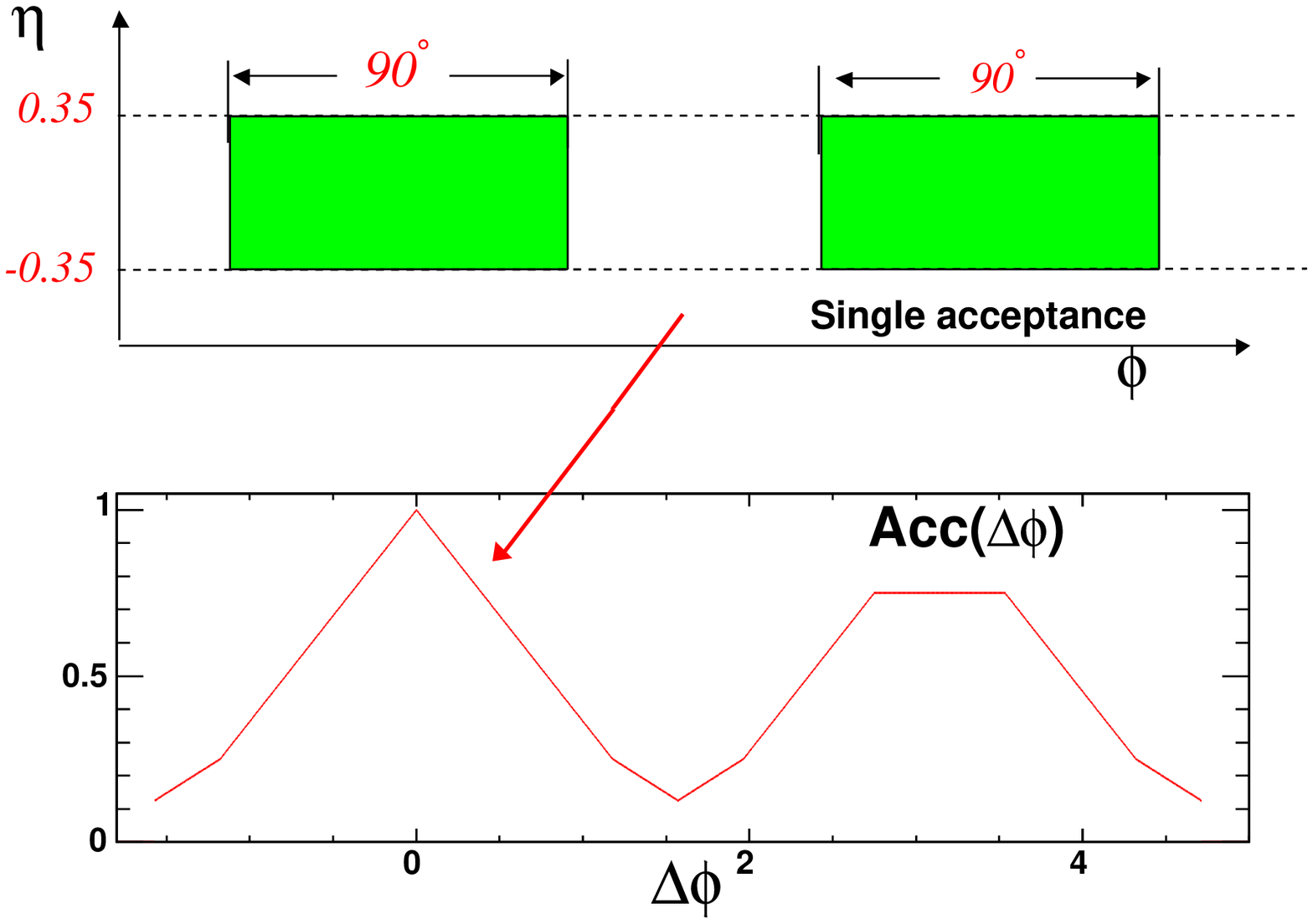,width=1.0\linewidth}
\end{flushright}
\end{minipage}
& \hspace*{-1.0cm}
\begin{minipage}{0.4\linewidth}
\begin{flushleft}
\begin{tabular}{l}
\vspace*{0.1cm}
\begin{minipage}{1.0\linewidth}
\begin{flushleft}
\epsfig{file=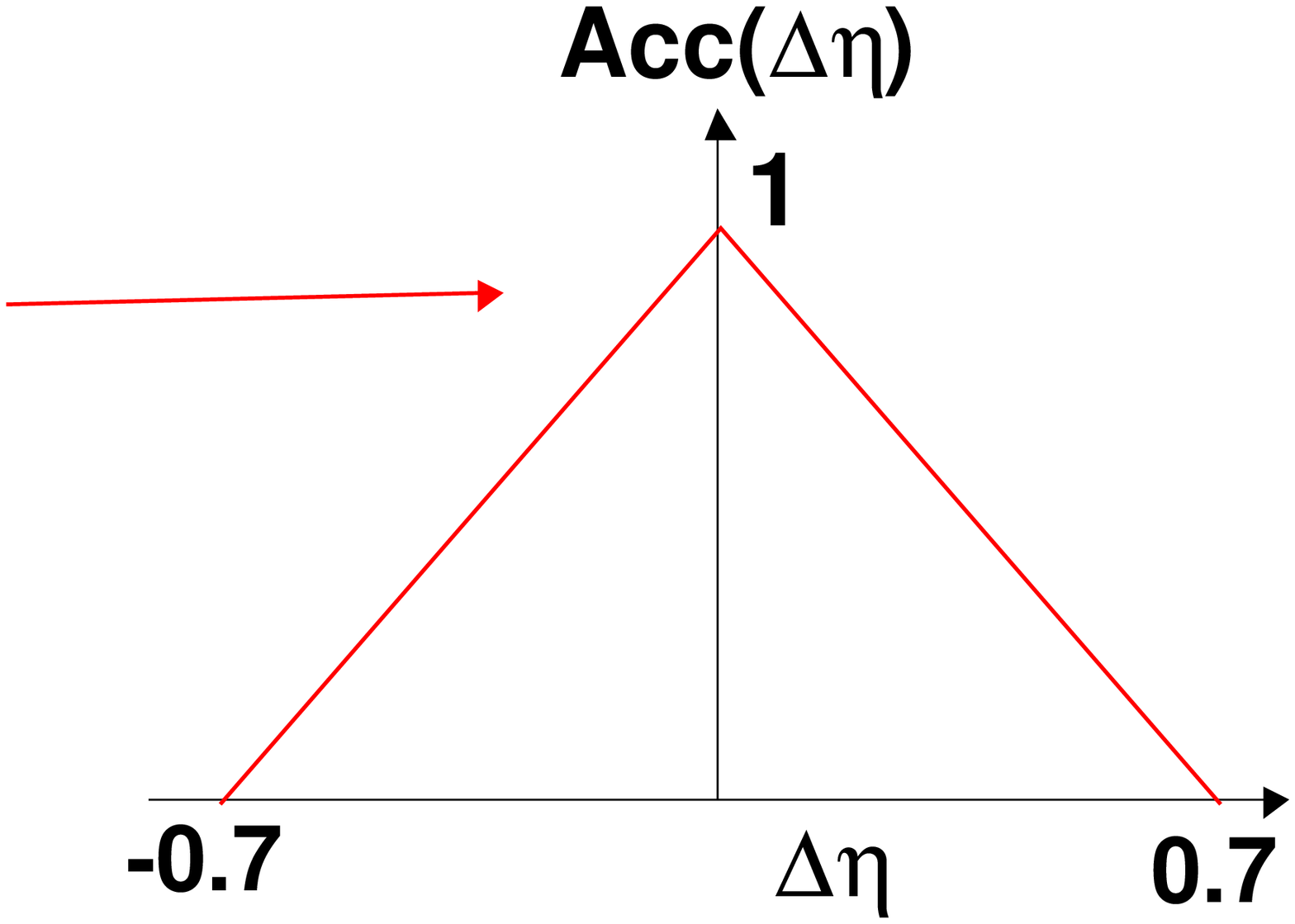,width=0.7\linewidth}
\end{flushleft}
\end{minipage}
\\
\hspace*{-2.5cm} \vspace*{1.0cm}
\begin{minipage}{1.3\linewidth}
\begin{flushleft}
\caption{\label{fig:acc} The PHENIX ideal single particle
acceptance (top left) and corresponding pair acceptance in
$\Delta\eta$ (top right) and $\Delta\phi$ (bottom).}
\end{flushleft}
\end{minipage}
\end{tabular}
\end{flushleft}
\end{minipage}
\end{tabular}
\end{figure}

The PHENIX ideal single particle acceptance is shown in
Fig.\ref{fig:acc}a :
$\phi\in[-33.75^\circ,56.25^\circ],[123.75^\circ,213.75^\circ]$
and $|\eta|<0.35$. The tracking efficiency has been assumed to be
100\%. The corresponding pair acceptance can be constructed by
convoluting the single acceptance for trigger and associated
particle and is shown in Fig.\ref{fig:acc}b for $\Delta\eta$ and
Fig.\ref{fig:acc}c for $\Delta\phi$. Interestingly, although the
PHENIX detector has only $\pi$ coverage in azimuth, the pair
acceptance actually is sensitive to the full range in
$\Delta\phi$, $\Omega_{\Delta\phi}=2\pi$; similarly the pair
coverage in $\Delta\eta$ is also doubled relative to single
particle $\eta$ acceptance, $\Omega_{\Delta\eta}=1.4$. The pair
phase space is $2\pi\times 1.4= 2.8\pi$, four times that for
single particle acceptance, $\pi\times 0.7= 0.7\pi$. However, this
increase is compensated by the decrease in overall pair
efficiency, which is only 25\%. There is only one point
($\Delta\phi=0,\Delta\eta=0$) where $Acc = 1$ (the probability of
detecting the associated particle when the trigger particle is
detected is 1).

The foreground distribution is modulated by the same pair
acceptance function:
\begin{eqnarray}
\label{eq:acc2} \frac{d^2N^{fg}}{d\Delta \phi d\Delta \eta }
&=&Acc(\Delta\phi,\Delta\eta)(\lambda m_0 +
jet(\Delta\phi,\Delta\eta))\\
&=& \lambda\frac{d^2N^{mix}} {d\Delta \phi d\Delta \eta }+
 \frac{d^2N^{jet}}{d\Delta \phi d\Delta \eta }.
\end{eqnarray}
%&=& \lambda{{d^2N^{mix} } \mathord{\left/
% {\vphantom {{d^2N^{mix} } {d\Delta \phi d\Delta \eta }}} \right.
% \kern-\nulldelimiterspace} {d\Delta \phi d\Delta \eta }}+
% {{d^2N^{jet} } \mathord{\left/
% {\vphantom {{d^2N^{jet} } {d\Delta \phi d\Delta \eta }}} \right.
% \kern-\nulldelimiterspace} {d\Delta \phi d\Delta \eta }}.

By dividing the foreground by the mix distribution, the $Acc$
cancels out, and we are left with a constant background plus the
jet signal. This ratio has correct jet shape, but the magnitude is
off by factor of $1/m_0$. Let's assume the original number of
trigger ($|\eta|<0.35$) and associated particles (within
$|\Delta\eta|<0.7$) per event are $n_{trig}$ and $n_{asso}$, and
those within PHENIX acceptance per event are $n\prime_{trig}$ and
$n\prime_{asso}$, respectively. Then the sum rule for original and
acceptance filtered mix event pair distribution are,
\begin{eqnarray}
\label{eq:acc3a} \int\int_{\Omega_{\Delta\phi}\times
\Omega_{\Delta\eta}} d\Delta\phi d\Delta\eta
\frac{d^2N^{mix}_0}{d\Delta\phi d\Delta\eta} =
N_{events}n_{trig}n_{asso} \quad,\\
\label{eq:acc3b}\int\int_{\Omega_{\Delta\phi}\times
\Omega_{\Delta\eta}} d\Delta\phi d\Delta\eta
\frac{d^2N^{mix}}{d\Delta\phi d\Delta\eta} =
N_{events}n\prime_{trig}n\prime_{asso}
\end{eqnarray}
where $N_{trig} = N_{events}n_{trig}$ is the total number of
triggers. From Eq.\ref{eq:acc2}-\ref{eq:acc3b}, we obtain the
following relation between the original(true) jet $CY$ and the
measured $CY$:
\begin{eqnarray}
\label{eq:acc4a} \frac{1}{N_a} \frac{d^2N^{jet}_0}{d\Delta\phi
d\Delta\eta}= \frac{1}{N\prime_a}\frac{1}{\epsilon_{asso}}
\frac{d^2N^{jet}/d\Delta\phi
d\Delta\eta}{\frac{\Omega_{\Delta\phi}\Omega_{\Delta\eta}
d^2N^{mix}/d\Delta\phi
d\Delta\eta}{\int\int_{\Omega_{\Delta\phi}\times
\Omega_{\Delta\eta}} d\Delta\phi d\Delta\eta
d^2N^{mix}/d\Delta\phi d\Delta\eta}},\\
\label{eq:acc4b} \epsilon_{asso} =
n\prime_{asso}/n_{asso}=\epsilon_{single}/\Omega_{\Delta\eta}
\end{eqnarray}
$\epsilon_{single}$ represents the single particle correction to
$2\pi$ in $\phi$ and 1 unit in $\eta$.

\subsection{$CY$ in $\Delta\phi$}

Building two dimensional correlation requires high statistics for
event mixing. Instead, correlation function is often built as
function of $\Delta\phi$ only, by integrating the 2D correlation
function over $\Delta\eta$. The following relations are true,
\begin{eqnarray}
\label{eq:cy1d1} \frac{1}{N_{trig}}\frac{dN_0^{jet}}{d\Delta\phi}
&=& \frac{1}{N_{trig}}\int_{\Omega_{\Delta\eta}}d\Delta\eta
\frac{d^2N_0^{jet}}{d\Delta\phi d\Delta\eta}\quad,\\
\label{eq:cy1d2}\frac{1}{N\prime_{trig}}\frac{dN^{jet}}{d\Delta\phi}
&=&
\frac{1}{N\prime_{trig}}\int_{\Omega_{\Delta\eta}}d\Delta\eta\frac{d^2N^{jet}}{d\Delta\phi
d\Delta\eta} \quad,\\
\label{eq:cy1d3}\textrm{and}\quad\frac{dN^{mix}}{d\Delta\phi} &=&
\int_{\Omega_{\Delta\eta}}d\Delta\eta\frac{d^2N^{mix}}{d\Delta\phi
d\Delta\eta}\quad.
\end{eqnarray}

To relate the measured $\Delta\phi$ distribution
(Eq.\ref{eq:cy1d2}) with the true $\Delta\phi$ distribution
(Eq.\ref{eq:cy1d1}), we require the following two assumptions:
\begin{itemize}
\item
The jet signal can be factorized in $\Delta\eta$ and $\Delta\phi$, i.e. $jet(\Delta\eta,\Delta\phi) = g_1(\Delta\phi)\times g_2(\Delta\eta)$\\
Typically, the jet signal can be approximated by,
\begin{eqnarray}
jet(\Delta\phi,\Delta\eta) = (C_1
e^{-\frac{\Delta\phi^2}{2\sigma_1^2}}+
C_3e^{-\frac{(\Delta\phi-\pi)^2}{2\sigma_3^2}})\times(C_2
e^{-\frac{\Delta\eta^2}{2\sigma_2^2}} +J(\Delta\eta))\quad.
\end{eqnarray}
In pp or pA collisions, near side jet widths in $\Delta\phi$ and
$\Delta\eta$ are the same, $\sigma_1=\sigma_2$. On the away side
the jet distribution in $\Delta\eta$ ($J(\Delta\eta)$) is
typically very wide due to the difference in momentum
fraction,$x$, of the two initial partons.

\item
The pair acceptance function can be factorized in $\Delta\phi$ and
$\Delta\eta$, i.e $Acc = Acc_1(\Delta\phi)\times
Acc_2(\Delta\eta)$. This condition is satisfied if the single
particle efficiency factorize in $\phi$ and $\eta$.
\end{itemize}
with these two assumptions, using Eq.\ref{eq:acc4a}, we can derive
following relation that connects Eq.\ref{eq:cy1d1} and
Eq.\ref{eq:cy1d2}.
\begin{eqnarray}
\label{eq:cy1d4}\frac{1}{N_{trig}}\frac{dN_0^{jet}}{d\Delta\phi}
=\frac{1}{N\prime_{trig}\epsilon_{asso}}\frac{dN^{jet}/d\Delta\phi}{\frac{\Omega_{\Delta\phi}dN^{mix}/d\Delta\phi}{\int_{\Omega_{\Delta\phi}}
d\Delta\phi dN^{mix}/d\Delta\phi}} \quad,\\
\label{eq:cy1d5} \epsilon_{asso} =
\frac{\int_{\Omega_{\Delta\eta}} d\Delta\eta
Acc_2(\Delta\eta)g_2(\Delta\eta)}{\int_{\Omega_{\Delta\eta}}
d\Delta\eta
Acc_2(\Delta\eta)}\frac{\epsilon_{single}}{\int_{\Omega_{\Delta\eta}}
d\Delta\eta g_2(\Delta\eta)}\quad.
\end{eqnarray}
Comparing with Eq.\ref{eq:acc4b}, we sees that Eq.\ref{eq:cy1d5}
have two more terms that are related to the $\Delta\eta$
correction. The first term is the pair acceptance weighted average
of jet signal over $\Omega_{\Delta\eta}$. The second term
($\int_{\Omega_{\Delta\eta}} d\Delta\eta g_2(\Delta\eta)$)
corresponds to the fact that only a certain fraction of jet signal
falls in the $\Omega_{\Delta\eta}$.

Eq.\ref{eq:acc4a} and Eq.\ref{eq:cy1d4} are equivalent: both
correct the jet yield to the true yield in $\Omega_{\Delta\eta}$,
i.e $|\Delta\eta|<0.7$. To account for the loss of jet yield
outside $\Omega_{\Delta\eta}$, an additional extrapolation factor
is needed. Because the jet shape in $\Delta\eta$ are quite
different for the same side and away side, we use two different
approaches. For the same side we assume the jet width are equal
for $\Delta\eta$ and $\Delta\phi$, i.e. $\sigma_1 =
\sigma_2$~\cite{deta}, and simply extrapolate $g_2$ (guassian) to
the full jet yield. On the away side, since the jet width in
$\Delta\eta$ is much broader than the typical PHENIX pair
acceptance, we assume $g_2$ is constant in $\Omega_{\Delta\eta}$
and no additional extrapolation is required. Under such approach,
Eq.\ref{eq:cy1d5} becomes,

\begin{eqnarray}
\hspace*{-2cm} \label{eq:cy1d6} \epsilon_{asso} =
\left\{\begin{array}{ll}
 \epsilon_{single}\frac{\int_{\Omega_{\Delta\eta}} d\Delta\eta
 Acc_2(\Delta\eta)\frac{1}{\sqrt{2\pi}\sigma_1}e^{-\Delta\eta^2/(2\sigma_1^2)}}{\int_{\Omega_{\Delta\eta}} d\Delta\eta
 Acc_2(\Delta\eta)}
 &\textrm{  full jet yield at the same side}
 \\
 \epsilon_{single}/\Omega_{\Delta\eta}&\textrm{ yield in $\Omega_{\Delta\eta}$ at the away side}
    \end{array}\right.
\end{eqnarray}
Since $Acc_2$ typically has a triangular shape, this correction is
easy to evaluate.
\section{Pythia Simulation}
\label{sec:simulation}

\begin{figure}[ht]
\begin{tabular}{cc}
\begin{minipage}{0.65\linewidth}
\begin{flushleft}
\epsfig{file=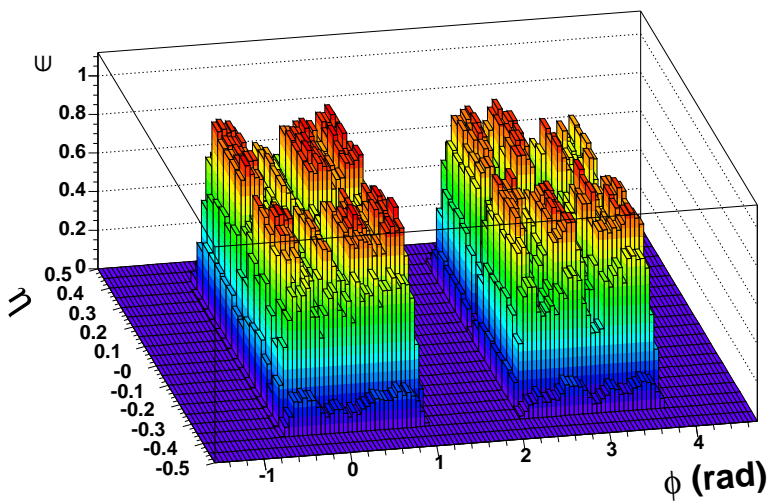,width=1.0\linewidth}
\end{flushleft}
\end{minipage}
& \hspace*{-2.4cm}
\begin{minipage}{0.4\linewidth}
\begin{flushleft}
\caption{\label{fig:cycheck1} A typical PHENIX single particle
acceptance/efficiency map for charged hadrons that was used in the
simulation and plotted as function of $\Delta\phi$ and
$\Delta\eta$.}
\end{flushleft}
\end{minipage}
\end{tabular}
\end{figure}

The extraction procedures for jet width and $CY$ are verified with
Pythia event generator~\cite{pythia}, using $\pi^{\pm}-h^{\pm}$
correlation. A single particle acceptance filter is imposed to
randomly accept charged particles according to the detector
efficiency. Fig.\ref{fig:cycheck1} shows the PHENIX style two
dimensional single particle acceptance filter used in the
simulation. The average efficiency in $2\pi$ in azimuth and 1 unit
of pseudo-rapidity is $\epsilon_{single} = 0.256$.

\begin{figure}[ht]
\begin{tabular}{cc}
\begin{minipage}{0.6\linewidth}
\begin{flushleft}
\epsfig{file=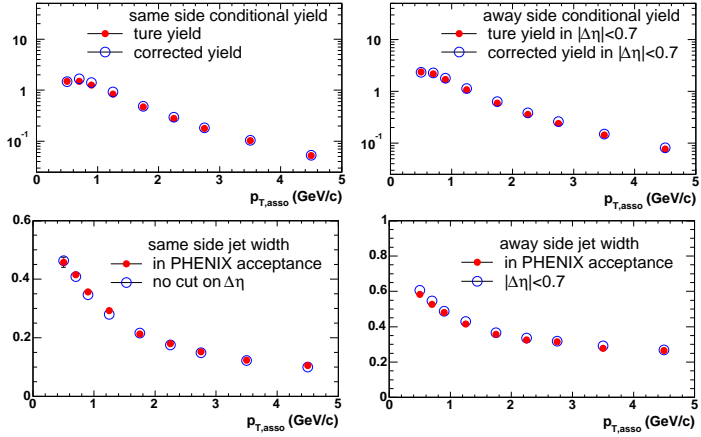,width=1.0\linewidth}
\end{flushleft}
\end{minipage}
& \hspace*{-2.5cm}
\begin{minipage}{0.5\linewidth}
\begin{flushleft}
\caption{\label{fig:cycheck2} The comparison of near side yield
(top left panel), near side width (bottom left panel), away side
yield(top right panel) and away side width (bottom right panel) as
function of $p_T$ of charged hadrons. These are obtained for
$\pi^{\pm}-h^{\pm}$ correlation from Pythia, with trigger pion
from 6-10 GeV/c. The open circles represent the quantities
calculated with the acceptance filter show in
Fig.\ref{fig:cycheck1}.}
\end{flushleft}
\end{minipage}
\end{tabular}
\end{figure}

We generated 1 million Pythia events, each required to have at
least one $>6$ GeV/c charged pions. To speed up the event
generation, a cut of $Q^2>100 GeV^2$ on the underlying
parton-parton scattering is required. These events were filtered
through the single acceptance filter. As an approximation, we
ignore the $p_T$ dependence of acceptance. The same event and
mixed pair $\Delta\phi$ distributions were then built by combining
the accepted $\pi^{\pm}$ and charged hadrons, where the trigger
$\pi^{\pm}$ is selected to be $6<p_{T,trig}<10$ GeV/c. The jet
width and raw yield were extracted by fitting the
$\frac{dN_{fg}}{\Delta\phi}/\frac{dN_{mix}}{\Delta\phi}$ with a
constant plus double gaussian function. The raw yields were then
corrected via Eq.\ref{eq:cy1d6} to full jet yield for same side
and to true yield in $|\Delta\eta|<0.7$ for away side. Meanwhile,
we also extract the true $CY$ and jet width without the acceptance
requirement. The comparison of the $CY$ and jet width with and
without the acceptance requirement are shown in
Fig.\ref{fig:cycheck2}. In the near side, the corrected yield (top
left panel) and width (bottom left panel) are compared with those
extracted without acceptance filter. In the away side, the yield
corrected back to $|\Delta\eta|<0.7$ (top right panel) and width
(bottom right panel) are compared with those extracted without
acceptance filter. The data requiring acceptance filter are always
indicated by the filled circles while the expected yield or width
are indicated with open circles.

\begin{figure}[ht]
\begin{tabular}{cc}
\begin{minipage}{0.6\linewidth}
\begin{flushleft}
\epsfig{file=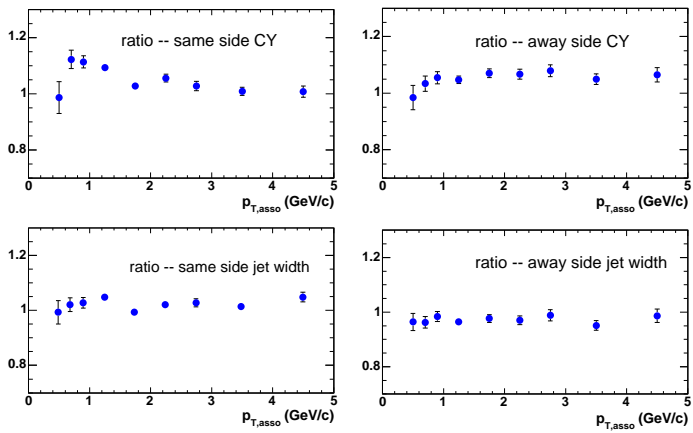,width=1.0\linewidth}
\end{flushleft}
\end{minipage}
& \hspace*{-2.0cm}
\begin{minipage}{0.45\linewidth}
\begin{flushleft}
\caption{\label{fig:cycheck3} The ratio of the jet width or
corrected yield obtained using event mixing method to those
obtained without acceptance filter.}
\end{flushleft}
\end{minipage}
\end{tabular}
\end{figure}

The agreement between the two data sets can be better seen by
plotting the ratios, which are shown in Fig.\ref{fig:cycheck3}.
The yields agree within 10\% and the widths agree within 5\%.
Since $\mean{j_{T_y}^2}$, $\mean{k_{T_y}^2}$ are derived from the
jet width(Eq.\ref{eq:jt1}-\ref{eq:kt2}), the agreement in width
naturally leads to the agreement in the $\mean{j_{T_y}^2}$,
$\mean{k_{T_y}^2}$. One can notice that there are some systematic
difference in the comparison of the yield at low $p_{T,asso}$.
This might indicate that the gaussian assumption is not good
enough when the jet width is wide and the extrapolation for
$|\Delta\eta|>0.7$ become sizeable (At $p_{T,asso}=0.5$ GeV/c, the
jet width $\sigma_{N} = 0.5$(rad), and the extrapolation is about
20\%.).

\section{Conclusion}
The formulae on $j_T$, $k_T$ and $CY$ are discussed in two
particle correlation framework. A more general definition of
$(j_{T_y})_{RMS}$ is found to be consistent with previous
approximation, but the $(k_{T_y})_{RMS}$ is lower by up to 10\%.
We have also demonstrated that the event mixing technique can
reproduce the jet width (thus the $j_T$ and $k_T$) for a limited
acceptance detector. With a correction factor that takes into
account the limited $\Delta\eta$ acceptance, we can also reproduce
the $CY$. Based on a Pythia simulation in PHENIX acceptance, our
procedure can reproduce the $CY$ within 10\% and the jet width
within 5\% at both the same side and away side.


\begin{thebibliography}{99}
\bibitem{pythiaeta} This was checked with pythia.
It varies by less than 5\% with in $|\Delta\eta|<0.7$.
\bibitem{deta} If $\sigma_1 \neq \sigma_2$,
$\sigma_2$ has to be measured independently. $\sigma_2$ can be
obtained by fitting the correlation function in $\Delta\eta$ with
a gaussian function. The correlation function in $\Delta\eta$ can
be build in the same way as the correlation function in
$\Delta\phi$.
\end{thebibliography}
\end{document}